\documentclass[conference]{IEEEtran}
\IEEEoverridecommandlockouts
\usepackage{cite}
\usepackage{amsmath,amssymb,amsfonts}
\usepackage{algorithmic}
\usepackage{graphicx}
\usepackage{textcomp}
\usepackage{xcolor}
\usepackage{booktabs}
\usepackage{multirow}
\usepackage{array} 
\usepackage[most]{tcolorbox} 
\usepackage[colorlinks=true, linkcolor=black, citecolor=black, urlcolor=black]{hyperref}

\definecolor{darkblue}{rgb}{0,0,0.8}

\tcbuselibrary{listings, breakable}

\def\BibTeX{{\rm B\kern-.05em{\sc i\kern-.025em b}\kern-.08em
    T\kern-.1667em\lower.7ex\hbox{E}\kern-.125emX}}
\begin{document}

\title{AutoPK: Leveraging LLMs and a Hybrid Similarity Metric for Advanced Retrieval of Pharmacokinetic Data from Complex Tables and Documents}


\author{Hossein Sholehrasa\textsuperscript{1,2}, 
Amirhossein Ghanaatian\textsuperscript{2}, 
Doina Caragea\textsuperscript{2}, 
Lisa A. Tell\textsuperscript{3},\\
Jim E. Riviere\textsuperscript{1},
Majid Jaberi-Douraki\textsuperscript{1,4,*}\thanks{*Corresponding Author: jaberi@k-state.edu}\\\\
\textsuperscript{1}1DATA Consortium and FARAD Program, Kansas State University, Olathe, KS, USA\\
\textsuperscript{2}Department of Computer Science, Kansas State University, Manhattan, KS, USA\\
\textsuperscript{3}FARAD, Department of Medicine and Epidemiology, University of California-Davis, Davis, CA, USA\\
\textsuperscript{4}Department of Mathematics, Kansas State University, Olathe, KS, USA\\
}



\maketitle

\begingroup
\renewcommand\thefootnote{}
\footnotetext{
© 2025 IEEE. Personal use of this material is permitted. Permission from IEEE must be obtained for all other uses.

This is the author's version of a paper published in: \textit{2025 IEEE 37th International Conference on Tools with Artificial Intelligence (ICTAI)}. The final published version is available at IEEE Xplore: \url{https://doi.org/10.1109/ICTAI66417.2025.00051}

}
\endgroup

\begin{abstract}
Pharmacokinetics (PK) plays a critical role in drug development and regulatory decision-making for human and veterinary medicine, directly affecting public health through drug safety and efficacy assessments. 
However, PK data are often embedded in complex, heterogeneous tables with variable structures and inconsistent terminologies, posing significant challenges for automated PK data retrieval and standardization.
AutoPK, a novel two-stage framework for accurate and scalable extraction of PK data from complex scientific tables. In the first stage, AutoPK identifies and extracts PK parameter variants using large language models (LLMs), a hybrid similarity metric, and LLM-based validation. The second stage filters relevant rows, converts the table into a key-value text format, and uses an LLM to reconstruct a standardized, machine-readable table.
Evaluated on a real-world dataset of 605 annotated PK tables, including captions and footnotes, AutoPK demonstrates significant improvements in precision and recall over direct LLM baselines. For instance, AutoPK with LLaMA 3.1-70B achieved an F1-score of 0.92 on half-life and 0.91 on clearance parameters, outperforming direct use of LLaMA 3.1-70B by margins of 0.10 and 0.21, respectively. Smaller models such as Gemma 3-27B and Phi 3-12B with AutoPK achieved 2–7 fold F1 gains over their direct use, with Gemma’s hallucination rates reduced from 60–95\% down to 8–14\%. Notably, AutoPK enabled open-source models like Gemma 3-27B to outperform commercial systems such as GPT-4o Mini on several PK parameters. AutoPK enables scalable and high-confidence PK data extraction, making it well-suited for critical applications in veterinary pharmacology, drug safety monitoring, and public health decision-making, while addressing heterogeneous table structures and terminology and demonstrating generalizability across key PK parameters.
Code and data are available at:
\href{https://github.com/hosseinsholehrasa/AutoPK}{\textcolor{darkblue}{https://github.com/hosseinsholehrasa/AutoPK}}

\end{abstract}


\section{Introduction}

Tabular data is a critical resource across various fields, from scientific research to business analytics. However, its inherent structural complexity, including schema variability, mixed data types, and implicit relationships, creates unique challenges for effective information retrieval and reasoning. Unlike free-form text, tables organize data into rows and columns, embedding structured relationships among textual, numerical, categorical, and logical elements. They often encode metadata in headers, hierarchies, footnotes, captions, and implicit references that provide an essential context for interpretation. This structured nature makes reasoning over tabular data challenging, particularly for cross-column or cross-row analysis tasks, which require advanced tools for understanding and processing such formats.
Natural Language Processing (NLP) models have been widely utilized for various tasks involving tabular data. These include table parsing for information extraction (TPIE) (e.g., OmniParser \cite{wan2024omniparser}, SPAGHETTI \cite{zhang-etal-2024-spaghetti}, and DiSCoMaT \cite{gupta-etal-2023-discomat}), table reconstruction (e.g., GTRNet \cite{luo2022gtrnet}), question answering (QA) (e.g., CABINET \cite{patnaik2024cabinet}, Dater \cite{Dater}, and GraphOTTER \cite{li-etal-2025-graphotter}), and table retrieval (e.g., TableRAG \cite{chen2024tablerag} or a similar approach \cite{chen2024table}). TPIE focuses on analyzing and interpreting structured data in tables to extract meaningful insights or specific information. This technique is used in data mining, document analysis, and automating the processing of complex datasets.

Despite significant advancements in NLP models and large language models (LLMs) to reason over free text data, their application to tabular data remains limited due to the structured and diverse nature of tables \cite{liu-etal-2024-rethinking}. Models like TAPAS \cite{herzig-etal-2020-tapas}
improved table understanding by incorporating tabular structures and relationships into training objectives. However, these models often require extensive fine-tuning and struggle to generalize across diverse or unseen table formats. In contrast, in-context learning approaches using LLMs 
offer greater flexibility by leveraging few-shot prompting to perform complex reasoning tasks without task-specific training. However, some challenges still persist, including structural variability, efficient cell lookup and reverse lookup, and the ability to filter out irrelevant or noisy data in tabular data types \cite{TablesMeetLLM}. Techniques such as chain-of-thought prompting showed promise in improving interpretability and reasoning, but still face limitations, particularly when handling large or highly complex tables \cite{wang2024chainoftableevolvingtablesreasoning}.

A particularly demanding real-world application of such tabular reasoning is found in the field of pharmacokinetics (PK), which studies how drugs are absorbed, distributed, metabolized, and excreted \cite{riviere2011comparative}. Key PK parameters, including half-life (HL; i.e., time for a drug concentration to reduce by half), clearance (CL; i.e., rate of drug elimination), maximum concentration (Cmax; i.e., highest drug concentration observed), time to maximum concentration (Tmax), mean residence time (MRT; i.e., average time a drug molecule remains in the body), and area under the concentration-time curve (AUC) are essential for predicting drug behavior and safety mandated by regulatory agencies \cite{PKparametersbook25}. These parameters play a vital role in determining optimal drug dosages, assessing potential toxicity, and complying with regulatory standards \cite{PKparametersbook25}. Importantly, these PK parameters serve as foundational variables from which other PK parameters can be derived \cite{bioavailability_calculation}.

Despite the importance of PK parameters, extracting them from the scientific literature remains highly challenging. PK data are often embedded in heterogeneous and irregular tables, with multi-header structures, merged cells, inconsistent formatting, and ambiguous abbreviations \cite{jaberi2021large}. The manual curation of PK data is highly time-consuming and prone to mistakes, and the accelerating growth of biomedical publications further limits the scalability of this manual curation \cite{smith2025automated}. While there has been progress in table parsing and LLM-based reasoning, there is limited research on the automated extraction of PK data from scientific studies. Existing LLM-based approaches, when applied directly to tables, struggle with structural complexity and inconsistent formatting \cite{TablesMeetLLM}, limiting the reliability and scalability of automated PK data extraction and leaving much of the work dependent on manual curation.

To overcome these limitations, we design AutoPK as a robust framework for automated PK data extraction. Specifically, the main contributions of this study are as follows: 
\begin{enumerate}
\item  We systematically identify and normalize PK parameter variants across heterogeneous table structures.\item  We reconstruct standardized, machine-readable tables that preserve key contextual information from tabular data. 
\item We benchmark AutoPK against strong LLM baselines on a real-world dataset. 
\end{enumerate}
This study demonstrates that our integrated approach can improve extraction accuracy and scalability, achieving high F1-scores across key PK parameters such as half-life and clearance while lowering hallucination rates compared to direct LLM usage.


\section{Problem Definition and Motivation}

PK parameters, such as elimination HL, CL, and AUC, are fundamental to understanding drug behavior in the body and are critical for ensuring drug safety and efficacy \cite{xu2024silico}. They influence therapeutic outcomes and prevent adverse drug events, which cost over \$42 billion in 2017 \cite{durand2024evaluating}. Accurate PK data retrieval and interpretation support drug research, dosing, and patient safety. Extraction errors can lead to incorrect dosing, unsafe withdrawal times in food animals \cite{riviere2017guide}, and harmful reactions, underscoring the need for systematic, reliable data extraction \cite{golmohammadi2025comprehensive}.
In veterinary medicine, PK parameters are vital for confirming food safety and guiding Extra-Label Drug Use (ELDU) under the Animal Medicinal Drug Use Clarification Act (AMDUCA) \cite{mi2025chemical}. Programs such as the Food Animal Residue Avoidance Databank (FARAD) \cite{farad} in the United States, gFARAD in Canada \cite{cgfarad}, and similar European initiatives rely heavily on accurate PK data to recommend appropriate drug withdrawal intervals and ensure public health. These programs serve as food safety resources that support veterinarians, producers, and regulators \cite{sholehrasa2025predictive}. Their work, used by thousands of stakeholders annually, relies on manually curated PK information extracted from published literature to make science-based recommendations for ELDU and contamination response. For example, our team at the FARAD program has manually performed this task for over 40 years \cite{ZAD2023113920}. This manual process is labor-intensive, error-prone, and increasingly unsustainable due to the growing volume of published data and rising cost of staff recruitment and data management \cite{smith2025automated}. Automation is therefore beneficial and critical for scaling PK data extraction and enabling timely, accurate, and reproducible recommendations that impact food safety for millions.

\section{AutoPK}
\begin{figure*}[!t]
    \vspace{-2em}
    \centering
    \includegraphics[width=\textwidth]{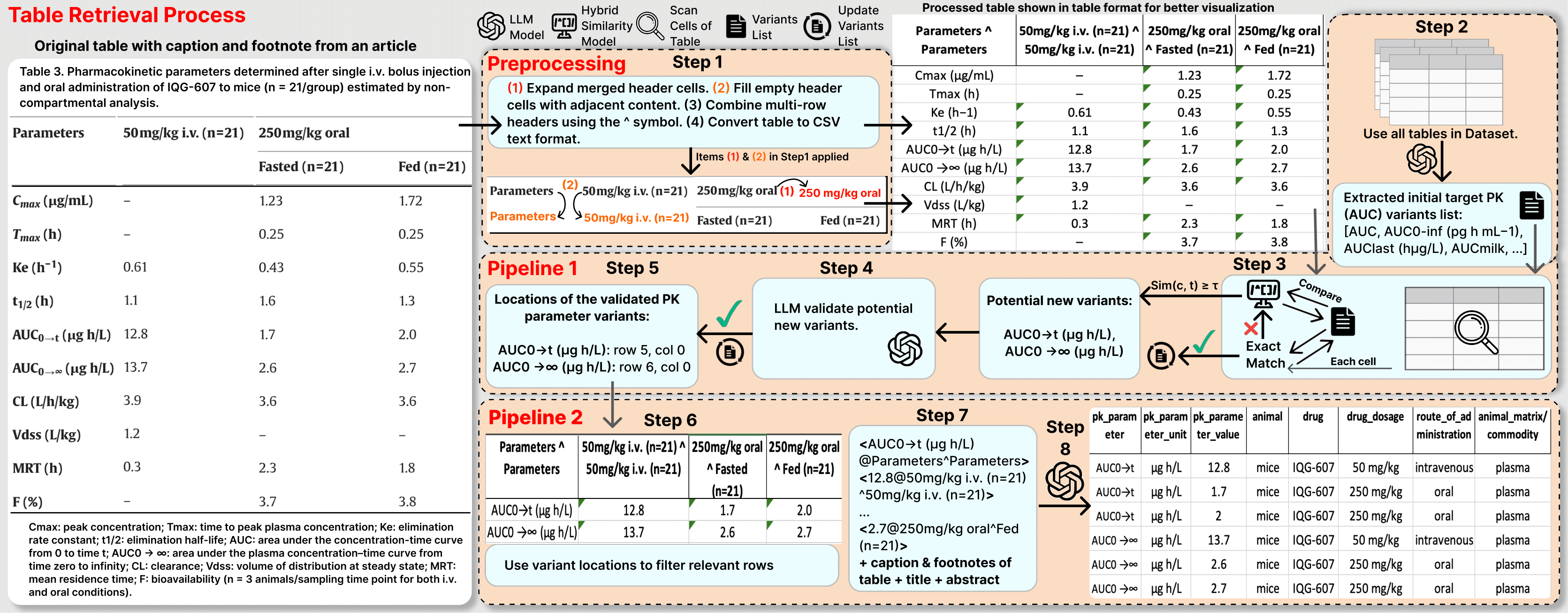}
    \vspace{-1.7em}
    \caption{Overview of the AutoPK table retrieval process. PK tables are preprocessed and scanned to extract initial AUC-related parameter variants. Expand the variants list via exact and hybrid similarity matching. Validated variants from an LLM are filtered and used to reconstruct structured CSV output.
}
    \label{fig:table_overview_process}
    \vspace{-1em}

\end{figure*}

To address the defined problem, we propose a framework, called AutoPK (consisting of a preprocessing step and two pipelines, Fig. \ref{fig:table_overview_process}), to identify key PK parameter variants and extract PK data primarily from tabular data in the scientific literature. While tabular data appears across various domains, its structural diversity necessitates preprocessing for automated data analysis. Our research focuses on the veterinary and biomedical domains, where accurate parameter extraction is essential for advancing research. Despite the apparent structure of tabular data in scientific literature, automatic extraction of PK data from these tables presents serious challenges. 
Tabular data often appears in various formats, each introducing unique challenges for automated interpretation. Single-header tables are the most straightforward, with each column labeled in a single row, making them relatively easy to process. In contrast, multi-header tables distribute header information across multiple rows, often inconsistently, leading to various layouts that complicate extraction. Another standard format is the block-structured table, where rows are grouped under high-level labels such as drug classes or species. These introduce hierarchical relationships that must be correctly identified to associate each row with its context. Additional structural challenges frequently arise across all of these types, such as merged header cells, empty cells, or column-wise headers, where parameter names are rotated vertically instead of listed horizontally. These features further obscure the relationships between labels and values. When such complexities occur together, especially in tables with inconsistent formatting or misaligned and inconsistent headers, they result in irregular tables that are difficult to interpret reliably, even for human readers. These irregular formats are not rare edge cases: in a pilot review of over 1000 PK-related studies, we found that more than 30\% contained non-standard table formats that required manual interpretation. This observation aligns with prior findings on the variability and lack of standardization in tabular data presentation across biomedical literature \cite{adams2022benchmarking}.

These structural complexities are further intensified by the limitations of LLMs in structured data understanding. While LLMs are powerful for text-based reasoning, they often fail in structured table understanding. Additionally, one other major challenge in table retrieval is posed by terminological inconsistencies in PK tables, as the same parameter may have multiple names and meanings across studies, e.g., half-life variation may lead to over a few hundred combinations, such as "T1/2," "HLgamma," or "Elimination Half-Life", etc. This lack of standardization creates a substantial barrier to reliable, scalable information retrieval using traditional parsing or even modern LLM-based methods. Pipeline 1 of AutoPK directly addresses this issue by systematically identifying these variations through a hybrid metric that combines similarity models and iterative LLM-based validation, ensuring that semantically and lexically equivalent variants are accurately recognized and mapped, regardless of their surface form. Subsequently, Pipeline 2 tackles the challenge posed by the structural diversity of tables. This pipeline minimizes hallucinations and information loss by filtering out large amounts of irrelevant or unnecessary information that may be present in the table. Rather than relying on rigid parsing rules, it transforms the filtered tables into a structured text representation that flattens complex layouts into a consistent, linear format. By converting each relevant cell into a key-value pair, Pipeline 2 creates a representation that abstracts away layout differences. This transformation enables text-based models, such as LLMs, to more effectively extract information from tabular data that would otherwise be difficult to parse due to structural variability.

In sum, automating PK data extraction from scientific tables is not just a technical convenience but a step toward enabling more efficient evidence-based workflows in veterinary pharmacology, with potential downstream benefits for public health.

\subsection{Preprocessing}
Before executing the AutoPK pipelines, we perform preprocessing of the tables (Fig. \ref{fig:table_overview_process}, Step 1). This standardizes the structure of each table. Since we work with tables where the header locations are known in advance, we focus specifically on normalizing the header rows. These headers often contain structural complexities such as merged cells, empty cells, or multiple rows. To handle merged cells, we duplicate their content across the full span of affected columns or rows, ensuring each column receives a complete and explicit label. Empty header cells are filled using content from adjacent cells to maintain consistency. When headers span multiple rows, we concatenate them by column, stacking the text from top to bottom within each column and joining them with the \^{} symbol. Then, we format the table as plain CSV text (using commas as separators) and feed it into AutoPK.

\subsection{Pipeline 1: PK Parameter Variant Identification}
We designed a multi-stage hybrid pipeline that combines LLMs with a similarity-based matching framework to address the challenge of identifying inconsistent PK parameter variants across structurally diverse and noisy tables. This approach allows progressively constructing parameter-specific variant sets without requiring manual rule design, extensive training, or reliance on ultra-large LLMs.

\subsubsection{Initial Extraction of Candidate PK Parameter Variants}
To initiate the process, we perform the following steps on all the tables from the dataset (Fig. \ref{fig:table_overview_process}, Step 2). Using the Gemma 3-27B language model, we prompted the system with a general instruction (Fig. \ref{fig:prompt-autopk-pipelines}, Prompt A) and five-shot examples related to the target PK parameter to extract the parameter variants. The examples are selected from annotated real instances. We selected partial representative cases that reflect how the target PK parameter variants may appear in actual data and used those to guide the model behavior. The output of this step is a list of possible candidate textual variants, including abbreviations, synonyms, and noisy expressions based on the dataset (e.g., ``HL alpha", ``T1/2Elm", etc.). These variants serve as a starting point for further expansion, with false-positive variants manually removed for improved performance of subsequent steps.



\subsubsection{Table-Wide Matching Using Hybrid Similarity Scoring}

We then iterated over every table in the dataset and compared each cell value against the list of candidate PK parameter variants (Fig. \ref{fig:table_overview_process}, Step 3). For efficiency, we first checked for an exact match (EM). If no EM was found, we computed a hybrid similarity score between the cell content and each candidate variant. For a given cell content $c$ and a candidate variant $v$, we computed a weighted similarity score $\text{Sim}(c, v)$ as:
\[
\text{Sim}(c, v) = \alpha \cdot \text{Cos}(c, v) + \beta \cdot \text{Lev}(c, v) + \gamma \cdot \text{Tok}(c, v)
\]
\noindent
where the final score $\text{Sim}(c, v)$ is computed as a weighted combination of three components: 1) cosine similarity ($\text{Cos}$) score, 2) normalized Levenshtein score ($\text{Lev}$), and 3) token overlap ($\text{Tok}$). The weights $\alpha$, $\beta$, and $\gamma$ are treated as hyperparameters and are fine-tuned based on validation set performance, as described below in the hyperparameter tuning section.

The three components of the similarity score, defined below,  collectively estimate the similarity between a cell string ($c$) and a target variant ($v$) across different criteria. 

\noindent
\textbf{Cosine Similarity} was used to evaluate the semantic relationship between the embeddings of $c$ and $v$. These embeddings were obtained using Bio+Clinical BERT \cite{bio_clinicalBERT}, a transformer-based language model that was first pretrained on biomedical literature via BioBERT \cite{BioBERT} and subsequently further pretrained on clinical notes from the MIMIC-III \cite{MIMIC-III} dataset. This model effectively captures domain-specific linguistic features relevant to both biomedical and clinical contexts. Cosine similarity was computed between the two embedding vectors, producing a score between 0 and 1, where 1 indicates high semantic similarity and 0 indicates no similarity.

\noindent
\textbf{Levenshtein Similarity} was incorporated to capture lexical similarity. The Levenshtein distance between a variant string $v$ and a cell string $c$ is defined as $LevDis(v, c)$ \cite{normalized_lev}.
The raw distance was normalized by the maximum length of the two strings and transformed into a similarity score as follows \cite{normalized_lev}:
\vspace{-0.5em}
\[
\text{Lev}(c, v) = 1 - \frac{LevDis(c, v)}{\max(|c|, |v|)}
\] \vspace{-0.5em}


\noindent
The Levenshtein similarity score ranges from 0 to 1, where 1 indicates an exact lexical match and 0 indicates no match.

\noindent
\textbf{Token Overlap} captures lexical similarity by quantifying the degree of overlap between the token sets of $c$ and $v$. Tokens are lowercased words extracted by splitting on whitespace, underscores, hyphens, and parentheses. We applied the Sørensen–Dice coefficient to compute this score \cite{sorensen_dice}:
\vspace{-0.2em}
\[
\text{Tok}(c, v) = \frac{2 \cdot |Tokens_c \cap Tokens_v|}{|Tokens_c| + |Tokens_v|}
\]
\vspace{-0.3em}
\noindent
where $|Tokens_c \cap Tokens_v|$ is the number of tokens shared between the two sets. The score ranges from 0 (no token overlap) to 1 (identical token sets). This measure captures partial matches between phrases that may differ slightly in formatting or word order.

If the overall hybrid similarity score, combining cosine similarity, normalized Levenshtein similarity, and token overlap, between a cell and any known variant exceeds a threshold of $\tau$, the cell is flagged as a \textit{potential new variant}.

\subsubsection{LLM-Based Validation of New Variants}

Any new variant that exceeds the similarity threshold is validated using Gemma 3-27B (Fig. \ref{fig:table_overview_process}, Step 4) to minimize false positives. A prompt is constructed to compare the new variant to examples of candidate PK parameter variants. If the LLM confirms the match, the new variant is added to the evolving list of candidate variants. This process ensures the candidate list improves as we iterate through the dataset while filtering out unrelated or misleading matches.

The final output of Pipeline 1 is a list of all identified variants of the target PK parameter, each paired with its location within the table (Fig. \ref{fig:table_overview_process}, Step 5). These variant-location pairs are then used in Pipeline 2 to extract the relevant data.

\subsection{Pipeline 2: Final Table Reconstruction}

\subsubsection{Table Simplification}

Once Pipeline 1 identifies variants of a specific PK parameter within a table and records their locations, Pipeline 2 uses these variant-location pairs to filter the relevant rows and construct the final table. For each matched instance, we retrieve its exact position and filter the corresponding row, while retaining the original table headers to preserve context. If any of the matched variants appear in column headers instead of row cells, we first transpose the table to enable consistent row-wise filtering. This approach isolates only the segments of the table directly related to the target PK parameter, discarding unrelated content. The result is a streamlined, filtered version of the table (Fig. \ref{fig:table_overview_process}, Step 6) focused exclusively on the selected parameter and its associated information. Since a single table may contain multiple surface forms of the same parameter (e.g., "HL $\alpha$" and "T1/2 eli" as variants of half-life), all matching rows corresponding to these variants are preserved.

\subsubsection{Table-to-Text Conversion and LLM-based Table Reconstruction}

After extracting the target rows, we proceeded to Step 7, where the simplified table was transformed into a structured text format (Fig. \ref{fig:table_overview_process}, Step 7). Each cell in the target rows was paired with its corresponding header using the "@" symbol, creating a consistent key-value representation for each PK parameter and its associated information (e.g. \texttt{<50gr@Dosage\^{}Chicken> <T1/2@Parameter\^{}Chicken>}
). These formatted entries, along with the table's caption and footnote, if available, were then passed to an LLM (Fig. \ref{fig:table_overview_process}, Step 8) using a 5-shot prompting \cite{chen2025benchmarking} (Fig. \ref{fig:prompt-autopk-pipelines}, Prompt B) approach to reconstruct the final output table in CSV format. Since PK parameters often span multiple rows, the LLM processed the entire set of formatted entries. It generated a complete table aligned with a predefined column structure specified in the prompt. Finally, we applied postprocessing to standardize the output tables, ensuring compatibility for meta-analysis and evaluation. To achieve this, both the generated tables (LLM outputs in CSV format) and the labeled ground-truth tables were normalized to share a consistent structure. This standardization step involved removing rows with empty PK parameter values, converting all text to lowercase, expanding abbreviations to full terms, and formatting numeric values uniformly to ensure reliable comparison and analysis.

\begin{figure}[!t]
\vspace{-1.8em}
\centering
\begin{tcolorbox}[
  enhanced,
  colback=gray!3!white,              
  colframe=gray!60!black,            
  coltitle=white,                    
  colbacktitle=gray!50!black,        
  fonttitle=\bfseries\small,         
  title=Prompts for AutoPK Pipeline 1 and Pipeline 2,
  boxrule=0.4pt,                     
  arc=4pt,                            
  top=6pt, bottom=6pt, left=6pt, right=6pt,  
  drop shadow={black!50!white},       
  width=\linewidth,
]
\small
\textbf{Prompt A – Variant Extraction (Pipeline 1-Step 1):}

\textbf{Input:} \texttt{\{Table in CSV Format\}} 

\textbf{Instruction:} Extract all variants of \texttt{\{pk parameter\}} (in various forms, like \texttt{\{variants Aliases\}}) based on the table provided. Write the exact names in the format of \texttt{<\$variant\$>} using \textnormal{\$\$} symbols like \textnormal{\$variant1\$}, \textnormal{\$variant2\$}, etc, without adding any extra text and without further information.  
Only provide \texttt{\{pk parameter\}} exactly as shown in the table without any changes. If a variant is embedded in a multi-header format like \textnormal{random1\^{}variant1\^{}random2}, return only what relates to \textnormal{variant1}, e.g., \textnormal{\$variant1\$}. It can be more than 1 form of \texttt{\{pk parameter\}} in the table. 
Do not include any forms where: \texttt{\{Non Variants Alias\}}  

\textbf{Answer format:} \textnormal{\$variant1\$,\$variant2\$}

\vspace{.2cm}
\hrule
\vspace{.3cm}




\textbf{Prompt B – PK Table Reconstruction (Pipeline 2):}
Extract PK data from any tables may appear inside a scientific document. Return one and only one comma-separated table with the header in the exact column order shown below—no commentary, no extra columns, no blank lines. My tables are in the specific text representation format which I combined my target row to the header with \@ sign for each cell and header can combine to other header with '\^{}' if the table is multi-header table. I want to convert this into a table format with the following columns (if not exists any data only left it with None): 
pk\_parameter, pk\_parameter\_unit, pk\_parameter\_value, animal, drug, drug\_dosage, route\_of\_administration, animal\_matrix/commodity

Extraction rules:
\texttt{\{Short Explanation About Each Column of Table\}}

\textbf{Inputs:}  
This is my custom format table: \texttt{\{Custom Format Table\}}  
This is footnote of my table in document: \texttt{\{Table Footnote\}}  
This is caption of my table in document: \texttt{\{Table Caption\}}  
This is title of my document: \texttt{\{Article Title\}}  
This is abstract of my document: \texttt{\{Article Abstract\}}

Output Produce nothing except the final CSV lines in the order specified.

\textbf{Answer format:} Table in CSV comma format text.

\end{tcolorbox}
\vspace{-0.5em}

\caption{The prompts used in AutoPK are summarized for (A) extracting PK parameter variants and (B) reconstructing CSV tables from simplified key-value text representations. Variables like \texttt{\{pk parameter\}}, \texttt{\{variants
Aliases\}}, and table metadata are dynamically replaced at runtime.}
\label{fig:prompt-autopk-pipelines}
\end{figure}

\subsection{Baseline: LLM Direct Table Retrieval}
We used various LLMs, including LLaMA 3.1-70B, Phi 3-12B, Gemma 3-27B, and GPT-4o Mini (accessed via API), to directly extract PK data from input tables in CSV format. Each model was prompted using a 5-shot approach, with examples designed to guide consistent extraction. Associated captions and footnotes were included, along with the title and abstract of articles when available. We also applied the same preprocessing (Fig. \ref{fig:table_overview_process}, Step 1) and postprocessing steps used in the AutoPK framework to ensure consistency across all outputs.


\section{Experimental Setup}

\subsection{Dataset}

We constructed a real-world dataset consisting of scientific tables and their surrounding textual context, including table captions and footnotes, as well as the title and abstract of the corresponding scientific articles. Table \ref{tab:dataset_stats} summarizes the key statistics of our datasets.

To build the real-world dataset, we used a publicly available PK-specific web crawler tool, as described in \cite{ramachandran2023automated}, to collect 1,088 XML-formatted full-text published articles. These articles include PK data from various species such as goats, chickens, rats, dogs, and cats. We extracted each XML file's article title, abstract, and full content of each table, including the table data, footnotes, and captions, by parsing the relevant XML tags. For table normalization, we processed multi-row headers by identifying rows in the \texttt{<thead>} tag and concatenating their contents column-wise, using the caret symbol (\textasciicircum) delimiter to preserve structure. Merged cells, as specified in the XML, were flattened by splitting them into individual cells and duplicating the original content across the full span to maintain layout consistency. This process yielded 1,522 tables that included PK data from 1,088 articles, with some articles containing multiple tables. Due to the potential for errors and the high cost of manual annotation, we selected and labeled a subset of 605 tables for assessing method performance, and used 5 for training/prompting, 180 (30\% of 600) for validation/hyperparameter fine-tuning, and 420 for testing.


\begin{table}[ht]
\vspace{-1.8em}

\centering
\caption{Summary statistics of the real-world dataset, including average number of rows and columns, table structure, and PK parameter variant counts.}
\begin{tabular}{>{\raggedright\arraybackslash}p{5.8cm} c}
\hline
\textbf{Statistic} & \textbf{Values} \\
\hline
\#Tables (train/val/test) & 5 / 180 / 420 \\
Avg \#rows/cols/multi-header-rows input tables & 8.63 / 5.43 / 2.35 \\
Avg \#rows/cols output tables & 21.56 / 8.00 \\
Unique HL / AUC / CL variants & 338 / 602 / 370 \\
Unique MRT / CMAX / TMAX variants & 61 / 161 / 74 \\
Single/multi-header/block-structured table types & 62\% / 26\% / 12\% \\
\hline
\end{tabular}
\label{tab:dataset_stats}
\vspace{-1.0em}

\end{table}

\subsection{Experimental Settings}
We conducted experiments using multiple LLMs, including GPT-4o Mini, LLaMA 3.1-70B, Phi 3-12B, and Gemma 3-27B. The GPT-4o Mini model was accessed via the OpenAI API, while the rest of the models were served through the LiteLLM API provided by the National Research Platform \cite{nrp}, running Python 3.12.4 and PyTorch 2.3.0.
All models used in AutoPK pipelines were prompted (Fig. \ref{fig:prompt-autopk-pipelines}) using a 5-shot format (training data subset). The temperature was set to 0.0 to ensure deterministic output, with a top-p value of 0.95 to allow for slight variability.

The expected output from each model is PK data in a CSV-formatted table with the following standardized schema:
\texttt{pk\_parameter, pk\_parameter\_unit, pk\_parameter\_value, animal, drug, drug\_dosage, route\_of\_administration, animal\_matrix/commodity}.

\subsection{Evaluation Protocol}
For each table in our dataset, labeled annotations specify the target PK parameter variants used to evaluate the first pipeline, as well as the corresponding PK data in table format, including dosage, route of administration, and other relevant information described in the Experimental Settings section.

\noindent
{\textbf{Pipeline 1:}}
For the evaluation of pipeline 1 of our framework, we compare the set of PK parameter variants identified by our method against the labeled ground-truth annotations using the EM approach. If a parameter variant extracted from the table matches a ground-truth label exactly (including formatting and spelling), it is considered a correct identification. This evaluation provides a direct measure of the precision and recall of the parameter variant identification component, ensuring that only precisely extracted variants are counted as correct matches in subsequent analyses.

\noindent
{\textbf{Pipeline 2:}}
To assess the accuracy of the final table reconstruction against the ground-truth, we implemented a cell-level evaluation consisting of two stages: (1) alignment of table structures and (2) comparison of corresponding cell values.

\noindent
\textbf{\it Table Alignment:} In the first stage, we align the structure of the generated table with the ground-truth table. Column alignment is performed by computing the Levenshtein similarity between column headers. A match is established if the normalized similarity score exceeds a threshold $\delta$. Any unmatched ground-truth table columns are treated as missing and are added to the generated table with empty values. Extra columns present in the generated table but not aligned to any ground-truth columns are tracked as extra columns. Once the columns are aligned, we proceed to row alignment. Each row in the ground-truth table is matched with the most similar row in the generated table based on the token overlap of cell contents. The average token overlap score is computed across all columns. A row is considered aligned if its similarity score exceeds a threshold $\theta$, and we choose the best score for row alignment. Unmatched rows in the generated table are considered extra rows, and unmatched rows in the ground-truth are treated as missing. After alignment, missing rows are inserted into the generated table with empty cells to match the ground-truth structure, ensuring both tables have identical shapes and order.

\noindent
\textbf{\it Cell Comparison:} In the second stage, we compare the aligned cell values between the ground-truth and generated tables. For each non-empty ground-truth cell, we compute Levenshtein similarity for numeric values and PK parameter variants, and cosine similarity for drug and animal entries, comparing ground-truth and generated cell values after normalization and string cleaning. If the similarity exceeds a threshold $\kappa$, the cell is considered a correct match (true positive). If the generated cell differs significantly from the ground-truth cell and the extra rows and columns (number of their cells) in the generated data, it is counted as a false positive. Missing cells in the generated table (i.e., NaN values for the generated table where the ground-truth has data) are counted as false negatives.

Finally, to quantify the performance of our table reconstruction method, we calculate standard evaluation metrics including precision, recall, and F1-score. Here, true positives represent the number of correctly matched cells, false positives include both incorrect cell values and hallucinated cells (from extra rows/columns), and false negatives correspond to missing cell values in the generated table.

\subsection{Hyperparameter fine-tuning}
We performed a validation set to fine-tune hyperparameters and select optimal settings. Specifically, the F1-scores were averaged across all PK parameters and used to determine the best configuration.

In the first step, we tested several LLMs to generate initial candidate variant lists. Based on validation performance, Gemma 3-27B achieved the highest overall F1-score of 0.86 across all PK parameters. In comparison, LLaMA 3.1-70B and Phi 3-12B trailed by 3 and 48 points, respectively, confirming Gemma's selection as the backbone for this stage.

To optimize similarity scoring, we tested about 900 combinations of weights and thresholds for all PK parameters. More than 130 configurations achieved an F1-score above 0.97 for each PK parameter. Among these, over 120 configurations generalized across parameters with an average F1-score above 0.97. The best-performing configuration with an average F1-score of 0.99 for similarity scoring has the following values: a threshold $\tau=0.69$ and weights $\alpha=0.6$, $\beta=0.2$, and $\gamma=0.2$ for cosine similarity, Levenshtein similarity, and token overlap, respectively. The presence of numerous high-performing configurations indicates that the similarity scoring metric is robust and not highly sensitive to specific weights and threshold configurations.

Based on empirical tests in the validation set, we selected threshold values that performed well in each stage of the table comparison process. For column alignment, we used a normalized Levenshtein similarity threshold of $\delta = 0.75$. For row alignment, an average token overlap score above $\theta = 0.5$ gave satisfactory results. At the cell level, a similarity threshold of $\kappa = 0.8$, applied using Levenshtein or cosine similarity depending on the data type, helped to distinguish the correct matches from the incorrect or missing values.

\section{Results \& Discussion}


%



\subsection{Comparative Evaluation of AutoPK}
We evaluated AutoPK on a test set of 420 samples across six key PK parameters: HL, AUC, CL, MRT, CMAX, and TMAX. As shown in Table~\ref{tab:method_comparison}, integrating LLMs into AutoPK’s framework dramatically boosted performance across all metrics. Most notably, AutoPK combined with LLaMA 3.1-70B achieved the strongest performance overall, with F1-scores above 0.89 on every PK parameter, including 0.95 on MRT, 0.91 on CL, and 0.92 on HL, CMAX, and TMAX. These results represent the strongest performance among all baselines tested in this study, showing how well powerful LLMs work together with the power of AutoPK's hybrid design. Even more impressive is how AutoPK turns weak-performing LLMs into high-performing systems. Models like Phi 3-12B and Gemma 3-27B often struggled with precision when used directly, resulting in low F1-scores performance typically below 0.45 and hallucination rates ranging 60\% to 95\%, where hallucination rate refers to the proportion of total predictions that are entirely simulated and not present in the original table. These models often extract irrelevant data unrelated to the PK parameter. For example, they confuse “CL” with “Creatinine Level” during extraction, leading to incorrect data capture. They also misinterpret noisy table structures, which results in failures like missing or wrong extraction of drug dosage values.

However, when wrapped with AutoPK’s framework, both recall and precision are dramatically enhanced. AutoPK(Phi 3-12B) achieved a 4 fold to 7 fold increase in F1 across most metrics, with HL improving from 0.12 to 0.48, and MRT from 0.13 to 0.54. Similarly, AutoPK(Gemma 3-27B) raised CMAX F1 from 0.46 to 0.90, and overperformed GPT-4o Mini in HL at 0.88. For Gemma 3-27B, hallucination rates dropped from 70\% to 60\% to around 8\% to 14\% after applying AutoPK. These gains validate AutoPK’s strength as it not only relies on model scale but also actively corrects underlying model weaknesses. These results also benefited the larger LLMs. While LLaMA 3.1-70B, on its own, already performed well, adding AutoPK led to consistent gains in precision, recall, and F1 across all parameters. This improvement is grounded in AutoPK’s architecture: it filters irrelevant content, simplifies table structures, and re-formats data into a more readable layout by LLMs. This significantly reduces hallucinations and boosts extraction quality.

Interestingly, even GPT-4o Mini, a high-performing commercial model, benefited from AutoPK. Its F1 scores ranged from 0.85 to 0.91 when used directly, but with AutoPK, results became more consistent across parameters. Recall improved, while precision dipped slightly, leading to F1-scores that were either better, roughly the same, or lower. This suggests that AutoPK brings out borderline cases that GPT might otherwise miss, trading a bit of precision for broader coverage. Crucially, AutoPK(Gemma 3-27B) outperforms GPT-4o Mini in most PK parameters. This highlights a primary strength of AutoPK, which is how AutoPK enables smaller or open-source models to match or surpass commercial models like GPT-4o Mini.

In addition to accuracy, AutoPK also demonstrates greater efficiency in pipeline 2 compared with direct LLM. By first filtering each table to retain only the most relevant rows, the pipeline reduces the amount of text passed to the LLM. This results in fewer tokens being processed, which lowers computational cost and shortens inference time, while still preserving the critical context for extraction.

This efficiency, combined with consistent performance across diverse PK parameters, from HL and CL to CMAX and TMAX, highlights AutoPK’s strong generalization capabilities in the PK parameter field. Rather than being tuned for a single parameter or pattern, AutoPK adapts to varying terminology, table structures, and contextual clues across the entire PK parameter space. This robustness confirms that the framework is practical in isolated cases and scalable and reliable for broad PK data extraction, without the need for heavy training or extensive fine-tuning.

With AutoPK delivering consistently high precision and recall across all major PK parameters, the framework enables trustworthy and scalable PK data extraction from published literature. In contexts such as withdrawal time determination, residue risk assessment, or therapeutic drug monitoring, AutoPK's extraction accuracy directly supports evidence-based decisions that protect public health. With its generalization across diverse table formats and terminologies, AutoPK is a robust solution for enabling automated, high-confidence PK data workflows in research and regulatory environments.

{
\small
\begin{table}[t]
\caption{precision (P), recall (R), and F1-score (F1) metrics (M) of test dataset (420 samples) for each method in PK parameter data extraction across real-world dataset}
\label{tab:method_comparison}
\centering
\begin{tabular}{
>{\raggedright\arraybackslash}p{2.4cm}  
>{\centering\arraybackslash}p{0.1cm}  
>{\centering\arraybackslash}p{0.3cm}  
>{\centering\arraybackslash}p{0.3cm}  
>{\centering\arraybackslash}p{0.3cm}  
>{\centering\arraybackslash}p{0.3cm}  
c  
c  
}
\toprule
\textbf{Method} & \textbf{M} & \textbf{HL} & \textbf{AUC} & \textbf{CL} & \textbf{MRT} & \textbf{CMAX} & \textbf{TMAX} \\
\#Tables &  & \small{418} & \small{335} & \small{361} & \small{138} & \small{245} & \small{189} \\

\midrule
\multirow{3}{*}{Phi 12B} 
    & P & 0.08 & 0.08 & 0.05 & 0.09 & 0.06 & 0.03 \\
    & R & 0.35 & 0.30 & 0.20 & 0.45 & 0.41 & 0.17 \\
    & F1 & 0.12 & 0.11 & 0.07 & 0.13 & 0.10 & 0.04  \\
\midrule
\multirow{3}{*}{AutoPK (Phi 12B)} 
    & P & 0.43 & 0.38 & 0.42 & 0.47 & 0.37 & 0.24 \\
    & R & 0.68 & 0.72 & 0.70 & 0.89 & 0.78 & 0.56 \\
    & F1 & 0.48 & 0.43 & 0.48 & 0.54 & 0.42 & 0.29  \\
\midrule
\midrule
\multirow{3}{*}{Gemma 27B} 
    & P & 0.36 & 0.27 & 0.30 & 0.36 & 0.27 & 0.34  \\
    & R & 0.98 & 0.98 & 0.97 & 0.98 & 0.98 & 0.98 \\
    & F1 & 0.46 & 0.39 & 0.40 & 0.45 & 0.36 & 0.43 \\
\midrule
\multirow{3}{*}{AutoPK (Gemma 27B)} 
    & P & 0.85 & 0.79 & 0.82 & 0.87 & 0.82 & 0.84 \\
    & R & 0.99 & 0.97 & 0.99 & 0.99 & 0.99 & 0.99  \\
    & F1 & 0.90 & 0.85 & 0.88 & 0.91 & 0.88 & 0.90  \\
\midrule
\midrule
\multirow{3}{*}{GPT-4o Mini} 
    & P & 0.86 & 0.87 & 0.82 & 0.90 & 0.86 & 0.84  \\
    & R & 0.94 & 0.89 & 0.94 & 0.95 & 0.94 & 0.93 \\
    & F1 & 0.88 & 0.86 & 0.85 & 0.91 & 0.88 & 0.86 \\
\midrule
\multirow{3}{*}{AutoPK (GPT-4o Mini)} 
    & P & 0.82 & 0.81 & 0.82 & 0.85 & 0.82 & 0.82 \\
    & R & 0.96 & 0.96 & 0.99 & 0.98 & 0.99 & 0.98 \\
    & F1 & 0.87 & 0.86 & 0.87 & 0.90 & 0.88 & 0.88  \\
\midrule
\midrule
\multirow{3}{*}{LLaMA 70B} 
    & P & 0.79 & 0.82 & 0.74 & 0.87 & 0.82 & 0.77 \\
    & R & 0.89 & 0.84 & 0.74 & 0.80 & 0.89 & 0.90 \\
    & F1 & 0.82 & 0.80 & 0.70 & 0.81 & 0.83 & 0.80 \\
\midrule
\multirow{3}{*}{AutoPK (LLaMA 70B)} 
    & P & 0.89 & 0.85 & 0.87 & 0.93 & 0.88 & 0.87 \\
    & R & 0.98 & 0.97 & 0.99 & 1.00 & 0.99 & 0.99 \\
    & F1 & \textbf{0.92} & \textbf{0.89} & \textbf{0.91} & \textbf{0.95} & \textbf{0.92} & \textbf{0.92} \\
\bottomrule
\end{tabular}
\vspace{-1em}
\end{table}
}




\subsection{Ablation Study}

We conducted an ablation study to evaluate the contribution of different components in the first pipeline, focusing on how each step impacts the accuracy of PK parameter variant extraction. We applied the optimal settings to the test split to assess the end-to-end effectiveness of the pipeline. All results use a fixed threshold of 0.69 and weighting of 0.6 (cosine), 0.2 (Levenshtein), and 0.2 (token overlap).

Table \ref{tab:ablation_pipeline1} shows the ablation study results for Pipeline 1, reporting the average F1-score, precision, and recall across all PK parameters using the optimal configuration. These metrics are used to assess the impact of individual pipeline components on the accuracy (precision), completeness (recall), and overall balance (F1-score) of variant extraction. Analysis of the results indicates that Step 2, which initiates variant identification, is foundational. Its effectiveness is highly dependent on the number of in-context examples (shots) and the volume of input data; reductions in either lead to a notable drop in overall performance. Step 3 enhances recall by expanding the variant list through hybrid similarity matching, outperforming exact matching alone. However, this expansion introduces false positives alongside true positives, lowering precision without further filtering. Step 4 addresses this issue by applying LLM-based validation to filter candidate variants. This step significantly improves precision while slightly enhancing recall. Its absence leads to a marked decline in overall performance, emphasizing its importance in ensuring reliable and accurate PK parameter extraction.

\begin{table}[t]
\centering
\caption{Ablation study results for real-world dataset. Each section reports the average F1-score (F1), precision (P), and recall (R) across all PK parameters, comparing different variants of the AutoPK Pipeline 1. Each component is also determined by its Steps (S) number.}
\vspace{-0.5em}
\label{tab:ablation_pipeline1}
\begin{tabular}{@{}lccc@{}}
\toprule
\textbf{Ablation Variant}                             & \textbf{P} & \textbf{R} & \textbf{F1} \\
\midrule
\textbf{Full AutoPK} (All Pipeline 1 Components)                          & 0.97 & 1.00      & 0.99   \\
\quad – w/o Hybrid Similarity (S3, EM only)      & 1.00   & 0.90   & 0.95   \\
\quad – w/o LLM variant Validation (S4)                       & 0.25    & 0.92   & 0.38   \\
\quad – 0-shot Prompting (S2)                              & 0.26        & 0.44       & 0.29   \\
\quad – 50\% data for initial extraction (S2)              & 0.88        & 0.92       & 0.90   \\
\bottomrule
\end{tabular}
\vspace{-1.5em}
\end{table}

\section{Limitations and Future Work}
A key limitation of the current approach lies in its handling of highly complex and irregular table structures where critical data, such as drug, dosage, or species, are not located in the same row as the PK parameter variants. In such cases, the filtering mechanism may fail to correctly associate all related information, especially when the layout is ambiguous or requires cross-row reasoning. These scenarios challenge the assumption that key information is co-located with the PK parameter variant and require more advanced, context-aware filtering mechanisms. Future work could explore multi-modal reasoning techniques that incorporate layout-aware models, visual table parsing, or graph-based representations to capture relationships across non-contiguous cells. Such methods would enable the system to reason beyond strict row-level associations and handle tables with more complex spatial and semantic layouts. At a broader level, an important direction for future work is extending AutoPK to handle less structured sources such as PDFs and OCR-derived tables. This would broaden its applicability to a wider range of PK studies and enable seamless integration into automated literature curation pipelines. Combining AutoPK with robust document layout analysis and table recognition methods could reduce the dependency on XML inputs, thus improving its external validity. Finally, while AutoPK has been primarily evaluated against direct LLM usage, future work should include benchmarking against state-of-the-art table understanding models such as TAPAS. In particular, comparing TAPAS with AutoPK integrated into the TAPAS would clarify how much additional value AutoPK provides to specialized table-focused architectures. Furthermore, directly comparing AutoPK (TAPAS) with AutoPK (LLaMA 70B), the current best-performing configuration, would offer valuable insight into the trade-offs between leveraging domain-specialized table models and large general-purpose LLMs within the AutoPK framework.

\section{Conclusions}
AutoPK offers a robust, scalable solution for high-accuracy extraction and standardization of PK data from complex scientific tables. By combining LLMs with a hybrid similarity scoring metric and LLM-based validation, AutoPK overcomes core challenges in PK data retrieval, including structural variability, inconsistent parameter terminology, and noisy table layouts. The framework consistently outperforms baseline LLMs across all major PK parameters, enabling smaller or open-source models to compete or surpass commercial systems like GPT-4o Mini. In addition to boosting F1-scores and reducing hallucinations, AutoPK's strong generalization across varied table formats and PK parameter fields makes it well-suited for real-world deployment without extensive fine-tuning of LLMs. This adaptability highlights its potential for wide adoption in regulatory, clinical, and research applications, particularly in veterinary pharmacology and public health.
\vspace{-0.4em}

\section*{Acknowledgment}
This work was supported by the USDA via the FARAD program (Award No.: 2022-41480-38135, 2023-41480-41034, 2024-41480-43679, and 2025-41480-45282) and its support for the 1DATA Consortium at Kansas State University.

Research reported in this publication was also partially supported by the Cognitive and Neurobiological Approaches to Plasticity (CNAP) Center of Biomedical Research Excellence (COBRE) of the National Institutes of Health under grant number P20GM113109. The content is solely the responsibility of the authors and does not necessarily represent the official views of the National Institutes of Health.

This work also utilized resources available through the National Research Platform (NRP) at the University of California, San Diego. NRP has been developed and is supported in part by funding from the National Science Foundation, through awards 1730158, 1540112, 1541349, 1826967, 2112167, 2100237, and 2120019, as well as additional funding from community partners.
\bibliographystyle{IEEEtran}
\bibliography{references}

\end{document}